\documentclass{webofc}
\usepackage[varg]{txfonts}   
\usepackage{bm}
\usepackage{amsmath}
\usepackage{graphicx}
\usepackage{float}
\usepackage[usenames,dvipsnames]{color}
\definecolor{darkblue}{RGB}{0,0,196}
\usepackage[colorlinks=true,linkcolor=darkblue,citecolor=darkblue,urlcolor=darkblue]{hyperref}
\usepackage{setspace}
\usepackage{footmisc}
\usepackage[makeroom]{cancel}
\usepackage[utf8]{inputenc}

\newcommand{\prevchangeflag}[1]{{ #1}}

\newcommand{\changeflag}[1]{{ #1}}

\def\be{\begin{equation}}
\def\ee{\end{equation}}
\def\ba{\begin{eqnarray}}
\def\ea{\end{eqnarray}}

\newcommand{\mbf}{\mathbf}
\newcommand{\mrm}{\mathrm}
\newcommand{\Tr}{\mrm{Tr}}

\newcommand{\cl}{{\mathrm{cl}}}

\newcommand{\fig}{Fig.~}

\newcommand{\res}{Refs.~}

\newcommand{\pToFigs}{.}

\begin{document}
\title{Long and short distance behavior of the imaginary part of the heavy-quark potential}
%
%

\author{\firstname{Kirill} \lastname{Boguslavski}\inst{1}\fnsep\thanks{\email{kirill.boguslavski@tuwien.ac.at}} \and
        \firstname{Babak} \lastname{Kasmaei}\inst{2}\fnsep\thanks{\email{kasmaei@hood.edu}} \and
        \firstname{Michael} \lastname{Strickland}\inst{3}\fnsep\thanks{\email{mstrick6@kent.edu}}
}

\institute{Institute for Theoretical Physics, Technische Universit\"at Wien, 1040 Vienna, Austria 
\and
           Department of Chemistry and Physics, Hood College, Frederick, MD 21701, United States 
\and
           Department of Physics, Kent State University, Kent, OH 44242, United States
          }

\abstract{%
  The imaginary part of the effective heavy-quark potential is related to the total in-medium decay width of heavy quark-antiquark bound states. We extract the static limit of this quantity using classical-statistical simulations of real-time Yang-Mills dynamics by measuring the temporal decay of Wilson loops. By performing the simulations on finer and larger lattices, we are able to show that the nonperturbative results follow the same form as the perturbative ones. For large quark-antiquark separations, we quantify the magnitude of the non-perturbative long-range corrections to the imaginary part of the heavy-quark potential. We present our results for a wide range of temperatures, lattice spacings, and lattice volumes. We also extract approximations for the short-distance behavior of the classical potential.
}
\maketitle
\section{Introduction}
\label{intro}

The suppression of heavy quark-antiquark bound states, such as bottomonium, is one of the key signatures of the Quark-Gluon Plasma (QGP). Due to the interactions of the heavy quark with the stochastic medium of the soft degrees of freedom, the effective heavy-quark potential possesses an imaginary part which is related to the total in-medium decay width of the heavy quark-antiquark bound states.  This imaginary part has been determined using direct quantum field theoretic or effective field theory calculations \cite{Laine:2006ns,Brambilla:2016wgg,Brambilla:2017zei}.

The imaginary part of the heavy-quark potential has been calculated based on high-temperature quantum chromodynamics (QCD) calculations in the hard thermal loop (HTL) limit \cite{Laine:2006ns,Dumitru:2007hy,Brambilla:2008cx,Burnier:2009yu,Dumitru:2009fy,Dumitru:2009ni,Margotta:2011ta,Guo:2018vwy}, using effective field theory (pNRQCD) \cite{PhysRevD.21.203,Lucha:1991vn,Brambilla:2004jw,Brambilla:2010xn}, finite-temperature lattice QCD \cite{Rothkopf:2009pk,Rothkopf:2011db,Burnier:2012az,Burnier:2013nla,Burnier:2015nsa,Burnier:2015tda,Burnier:2016mxc,Burnier:2016kqm,Bala:2019cqu,Bala:2020tdt}, and real-time classical-statistical solutions of Yang-Mills theory in classical thermal equilibrium
\cite{Laine:2007qy,Lehmann:2020fjt}. In Ref.~\cite{Laine:2007qy} the authors presented first results for the imaginary part of the heavy-quark potential using classical-statistical Yang-Mills simulations on spatially 3D lattices of size $12^3$ and $16^3$.  

We extend the previous results to larger lattices up to $252^3$
and consider SU(2) and SU(3) gauge theories (see Ref.~\cite{Boguslavski:2020bxt} for details on our results).  Due to the use of rather large lattice sizes, we can now compute the imaginary part of the heavy quark potential at larger values of $r/a$ and reconstruct the functional form of the imaginary part of the heavy-quark potential for a much wider range of distances.

Herein we will present the results for the imaginary part of the heavy-quark potential obtained using classical Yang Mills (CYM) simulations of a thermalized gluonic plasma.  The use of CYM simulations is motivated by the fact that in situations where (a) gluonic occupation numbers are large, such as in thermal equilibrium for sufficiently low momenta or in the initial stages of heavy-ion collisions, and (b) the gauge coupling is weak $g^2 \ll 1$, vacuum contributions to observables are suppressed by powers of the gauge coupling. 

 The system is initialized close to thermal equilibrium using momentum-space initialization and the fields self-thermalize in real time before the extraction of the observables. 
The advantage of this procedure is that we can run simulations on very large lattices with small lattice spacings $a$ at moderate computational cost. 

An issue which can arise when dealing with classical-statistical treatments of gauge theories is that a finite ultraviolet limit does not exist due to the Rayleigh-Jeans divergence \cite{Kajantie:1993ag,Ambjorn:1995xm,Arnold:1996dy,Moore:1999fs,Berges:2013lsa,Epelbaum:2014yja}.  For this reason, it is important to identify a suitable manner in which one can scale results in order to extract relevant information. We will use the Debye mass $m_D$ computed in the hard-classical-loops framework to demonstrate that, when plotted as a function of $m_D r$, the imaginary part of the heavy-quark potential {is only mildly sensitive to the lattice spacing, or more generally to the simulation parameter $\beta \propto 1 / (g^2Ta)$, at small distances $m_D r \lesssim 1$.}

\subsection{Theory and numerical setup}
\label{sec-2}
We mainly consider the pure SU(3) gauge theory with the Yang-Mills classical action. We use a standard real-time lattice discretization approach where fields are discretized on cubic lattices with $N^3$ sites and lattice spacing $a$ (see, e.g., \res\cite{Boguslavski:2018beu,Berges:2013fga} and references therein for more details). In this real-time approach, spatial gauge fields are replaced by gauge links $U_j(t,\mbf x) \approx \exp\left( i g a A_j(t,\mbf x) \right)$ at discrete coordinates $x_k = n_k a$ for $n_k = 0, \ldots, N-1$, while temporal gauge with $A_0 = 0$ is used. 

We are interested in extracting the imaginary part of the classical potential $V_\cl(t,r)$ with $r \equiv |\mbf x|$. Following \res\cite{Laine:2006ns,Laine:2007qy}, it can be calculated using
\begin{align}
 i\partial_t C_\cl(t,r) = V_\cl(t,r) C_\cl(t,r) \, ,
\end{align}
as the asymptotic temporal slope of $\log[C_\cl(t,r)]$. The classical thermal Wilson loop  $C_\cl(t,r)$ is defined as
\begin{align}
 C_\cl(t,r) \equiv \frac{1}{N_c}\,\Tr \left\langle W[(t_0,\mbf x);(t,\mbf x)]\, W[(t,\mbf x);(t,\mbf 0)]\, W[(t,\mbf 0);(t_0,\mbf 0)]\, W[(t_0,\mbf 0);(t_0,\mbf x)] \right\rangle,
\end{align}
with temporal Wilson lines $W[(t_0,\mbf x);(t,\mbf x)] = \textbf{1}$ and spatial Wilson lines $W[(t,\mbf 0);(t,\mbf x)] = U_j(t,\mbf 0)U_j(t,\mbf{a}_j)U_j(t,2\,\mbf{a}_j) \cdots U_j(t,\mbf x)$ for $\mbf x = \mbf{\hat a}_j\,r$ and $\mbf{\hat a}_j = \mbf{a}_j/a$ being a spatial unit vector. Since the classical thermal state is homogeneous, the Wilson loop is additionally averaged over all lattice points by averaging over the reference coordinates $\mbf 0$. 

In order to extract the imaginary part of $V_\cl(t,r)$ we compute the time-dependence of $C_\cl(t,r)$.  Due to the imaginary part of the in-medium heavy quark potential, this quantity will decay exponentially at late times with the rate of exponential decay set by the imaginary part of $V_\cl(t,r)$. We then define the imaginary part of the static classical potential as the late-time limit 
\begin{align}
 {\rm Im}[V_{\rm cl}(r)] \equiv \lim_{t \to \infty} {\rm Im}[V_{\rm cl}(t,r)]\,.
\end{align}
We initialize the fields in a quasi-thermal configuration in momentum-space, and then allow them to self-thermalize dynamically.

In Ref.~\cite{Laine:2006ns}, an expression is given for the imaginary part of the heavy-quark potential to leading-order in the strong coupling constant using the continuum hard thermal loop framework and dimensional regularization.  The final result could be expressed compactly as
\be
{\rm Im}[V^{(2)}(r)] = - \frac{C_F g^2 T}{4\pi} \phi\left(m_{D}^{\text{HTL}}\, r\right) \, .
\label{htlimv}
\ee
where $C_F = (N_c^2 -1)/2N_c\,$, 
\be 
\phi\left( x \right) \equiv 2 \int_0^{\infty} {\rm d}z \frac{z}{\left(1+z^2 \right)^2}\left[ 1- \frac{\sin\left(zx\right)}{zx}\right] ,
\label{eq:phidef}
\ee
and $m_{D}^{\text{HTL}}$ is the continuum hard thermal loop Deybe mass. 

The result for the imaginary part of the classical potential in the infinite volume and infinite time limit ($N \rightarrow \infty,\ t \rightarrow \infty$) from second-order perturbation theory regularized on a cubic lattice (HCL: Hard Classical Loop) of size $\left(aN\right)^3$  is calculated in Ref.~\cite{Laine:2007qy}.
We find that the numerical HCL curves are fit very well by the functional form
\be 
{\rm Im}[V^{(2)}_{\rm cl}(r)] = g^2T A_{\infty}\phi\left(B\,m_D r\right) .
\label{Aphi}
\ee
with $\phi(x)$ given in Eq.~\eqref{eq:phidef}. 

\section{Results}
\label{sec-3-results}

\begin{figure}[h]
\centerline{
\includegraphics[width=0.8\linewidth]{\pToFigs/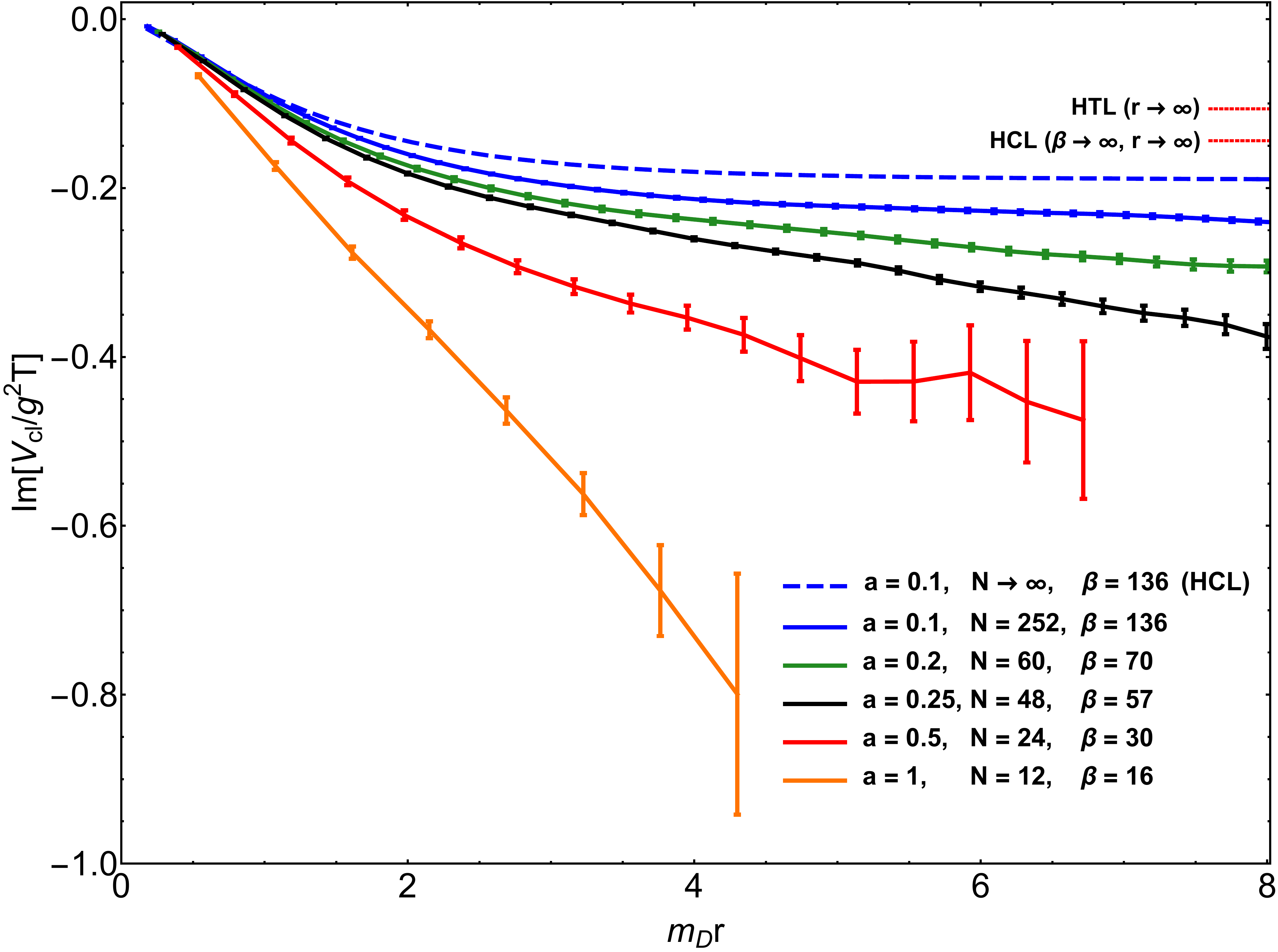}
}
\caption{ ${\rm Im}[V_{\rm cl}(r)]/g^2T$ for different lattice spacing from nonperturbative Wilson loops (solid lines) and from HCL perturbation theory for $a=0.1$ (blue dashed line)} 
\label{plot:imv_mDr}
\end{figure}

We performed simulations with $g^2 T_0 = 0.44$ and $a = \{1,\ 0.5,\ 0.25,\ 0.2\}$ keeping $aN = 12$ fixed and also for $g^2 T_0 = 0.45$ and $a =0.1$ with $N = 252$. 
The $a = 0.1$ CYM lattice simulation results are compared with the corresponding results (same $a$ and $T$) obtained from HCL perturbation theory  (included as a blue dashed curve) in \fig\ref{plot:imv_mDr}. For both our CYM lattice results and the HCL results, 
${\rm Im}[V_{\rm cl}(r)]/g^2T$ is plotted as a function of $m_D r$,
where we use the leading-order Debye mass $m_{D}^{\text{HCL}}$ calculated in HCL theory. 
On the right edge of the figure we indicate the asymptotic $r \to \infty$ values obtained using both the dimensionally-regularized continuum HTL result \eqref{htlimv} and the extrapolated $N,\beta \rightarrow \infty$ HCL result.

As can be seen in \fig\ref{plot:imv_mDr}, the CYM lattice results seem to \prevchangeflag{approach a finite large-$\beta$ form}
when plotted as a function of $m_D r$, with the CYM results with $\beta=136$ even overlapping with the corresponding HCL curve at $m_D r \lesssim 1$.
At large values of $r$ our CYM simulation results approach the corresponding HCL curve for the respective $\beta$ value. However, \fig\ref{plot:imv_mDr} indicates that \changeflag{with increasing $\beta$, the CYM potential is approaching the perturbative HCL large-$\beta$ limit from below.}

It should be noted that, while previous studies obtained the potential for different $\beta$ values as functions of $r/a$, we found that plotting it as a function of $m_D r$ makes the comparison more intuitive. Most importantly, this allows to \changeflag{study the classical potential even at large $\beta$ values while incorporating the dominant UV divergence into the mass.}%

\begin{figure}[t]
\centerline{
\includegraphics[width=0.48\linewidth]{\pToFigs/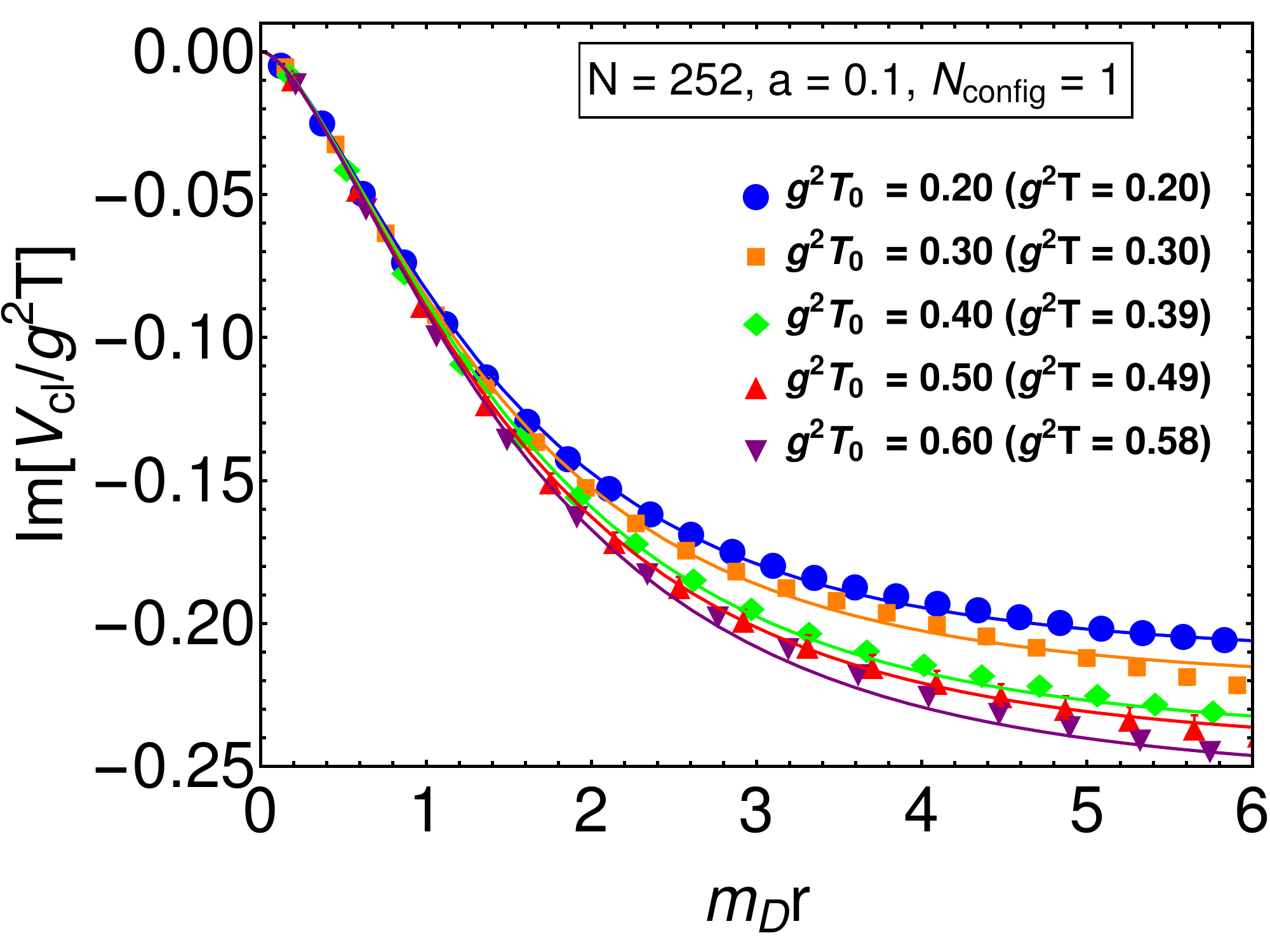}
$\;\;\;$
\includegraphics[width=0.48\linewidth]{\pToFigs/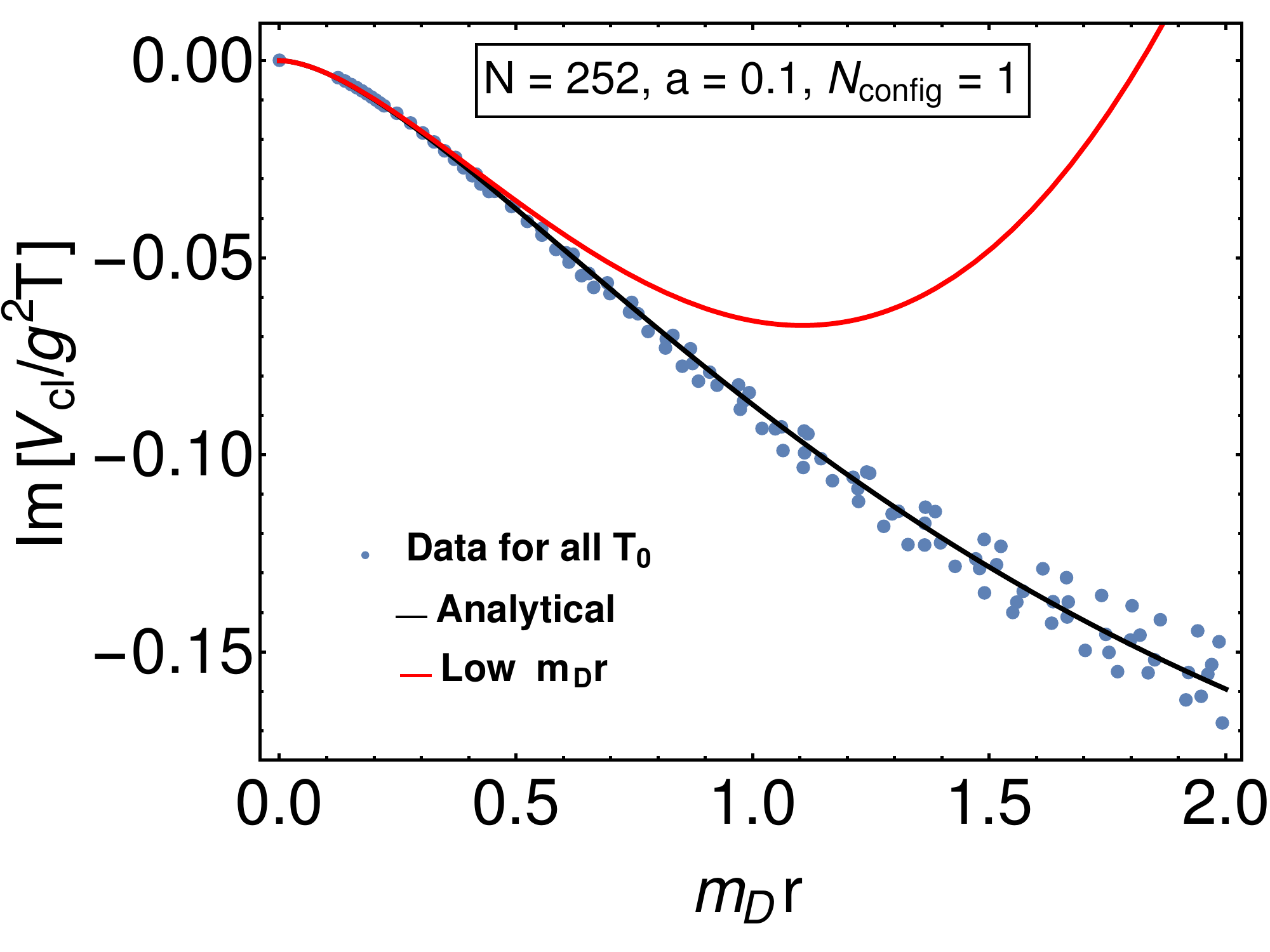}
}
\caption{
(Left) The extracted classical potential ${\rm Im}[V_{\rm cl}(r)]/g^2T$ as a function of $m_Dr$ for fixed lattice spacing $a = 0.1$ but for different temperatures.
(Right) The combined data of the left panel, with additional data sets is compared to the analytical form \eqref{Aphi}. 
}
\label{plot:short-dist}
\end{figure}

We \changeflag{provide evidence} that, at small distances, the classical potential extracted from our lattice simulation data also follows the functional form \eqref{Aphi} \changeflag{for different values of $\beta$}. This is shown in the left panel of \fig\ref{plot:short-dist}, where ${\rm Im}[V_{\rm cl}(r)]/g^2T$ for fixed lattice spacing $a=0.1$ is plotted
for different temperatures corresponding to $100 \lesssim \beta \leq 300$, as a function of $m_D\, r$. Fits to each data set using \eqref{Aphi} in the considered interval $m_D r \leq 6$ are included as continuous lines.

We can also perform a short-distance expansion of the fitting function for the imaginary part of the potential \eqref{Aphi}, neglecting terms $\mathcal{O}((m_D r)^4)$,
\begin{align}
 \label{eq:ImV2_lowX}
 {\rm Im}[V_{\rm cl}(r)] \simeq -r^2\,\frac{1}{9}\,|A_\infty|B^2\, g^2T\,m_D^2\left( 4-3\gamma - 3\log (B\, m_D\, r) \right).
\end{align}
\changeflag{The resulting curve} is shown as a red line in the right panel of \fig\ref{plot:short-dist} and is observed to agree well with our data points for $m_D\, r \lesssim 0.5$. 
\changeflag{Thus, for a wide range of $\beta$ values, the short-distance behavior of our CYM lattice data agrees well with the perturbative functional form \eqref{Aphi} and its leading short-distance expansion, which is parametrically given by $\left|{\rm Im}[V_{\rm cl}(r)]\right| \sim C_F\, g^2T\,\left(m_D\,r\right)^2\,\log (m_D\,r)$.}

\section{Conclusions}
\label{conclusion}

We used classical-statistical lattice simulations of the pure Yang-Mills theory to extract the imaginary part of the heavy-quark potential as a function of the quark-antiquark separation.  To carry out our simulations on large and fine lattices, we used a self-thermalization scheme to generate thermalized gauge field configurations which relied on initialization of chromo-electric fields in momentum-space followed by a period of self-thermalization.

We extended the previous classical-statistical lattice calculations of ${\rm Im}[V_{\rm cl}]$ by considering rather large lattice sizes. We also found that both the lattice simulation and the HCL results were very well approximated by a functional form which can be obtained from a leading-order hard-thermal loop calculation. Using fits of this form and then expanding the result at small $m_D r$, we were able to extract \changeflag{small-distance approximations for the imaginary part of the heavy-quark potential}. More details about our study can be found in Ref.~\cite{Boguslavski:2020bxt}.


\bibliography{ImV.bib}
%

%
%
%

\end{document}